\documentclass[a4paper,times]{article}
\usepackage{fullpage}
\usepackage{amsfonts}
\usepackage[utf8]{inputenc}
\usepackage[english]{babel}
\usepackage{algorithm}
\usepackage[noend]{algpseudocode}
\usepackage{graphicx}
\usepackage{pgfplots}
\usepackage{subcaption}
\usepackage{hyperref}
\usepackage[export]{adjustbox}
\usepackage{float}
\usepackage{tabto}
\usepackage{todonotes}
\usepackage{tabularx}
\usepackage{comment}
\usepackage{balance}

\usepackage[
backend=biber,
style=ACM-Reference-Format
]{biblatex}

\algrenewcommand\algorithmicindent{1.0em}%

\newcommand{\PKache}{\emph{PKache}}
\newcommand{\NetCache}{\emph{NetCache}}
\newcommand{\PFOUR}{\emph{P4}}
\newcommand{\TinyLFU}{\emph{W-TinyLFU}}
\newcommand{\KeysSet}{\emph{keys\_set}}
\newcommand{\CacheSet}{\emph{cache\_set}}
\newcommand{\SCN}{\emph{SCN}}
\newcommand{\WindowCache}{\emph{window cache}}
\newcommand{\MainCache}{\emph{main cache}}

\newcommand{\VLDBVERSION}{}

\begin{document}
\title{Limited Associativity Caching in the Data Plane}

\author{Roy Friedman \ \ \ \ Or Goaz \ \ \ \ Dor Hovav\\
Computer Science Department\\
Technion}

\maketitle
\begin{abstract}
In-network caching promises to improve the performance of networked and edge applications as it shortens the paths data need to travel.
This is by storing so-called hot items in the network switches on-route between clients who access the data and the storage servers who maintain it.
Since the data flows through those switches in any case, it is natural to cache hot items~there.


Most software-managed caches treat the cache as a fully associative region.
Alas, a fully associative design seems to be at odds with programmable switches' goal of handling packets in a short bounded amount of time, as well as their restricted programming model.
In this work, we present \PKache{}, a generic limited associativity cache implementation in the programmable switches' domain-specific \PFOUR{} language, and demonstrate its utility by realizing multiple popular cache management~schemes.

\end{abstract}


\pagestyle{plain}

\section{Introduction}
\ifdefined\VLDBVERSION
Caching is a fundamental technique for boosting systems performance.
In particular, software-managed caches, aka software caches, are employed in multiple data-stores and databases~\cite{AABMSY19,cascade,RistrettoProject,Ferdinand,Quaestor,CADS19,SlimDB,redis,Sundial,WLC20}, operating systems, middleware, streaming services, and is a major capability of edge computing.
The common motivation behind caching is to store data closer to the application than its source and avoid recalculating queries, query plans, and temporal indices.
For example, DRAM memory is faster and closer in the memory hierarchy than secondary storage, local storage is closer than accessing data over the Internet, etc.
This way, the requested information can be served quickly whenever the requested data is already in the cache, also known as a \emph{cache hit}.
In an edge computing setup, caching also saves network bandwidth and reduces the load on servers.
This is because when the data is served from a local edge cache, there is no need to load it from the remote~server~\cite{Ferdinand,Quaestor}.
\else
Caching is a fundamental technique for boosting systems performance.
In particular, software-managed caches, aka software caches, are employed in a wide range of systems and applications like operating systems, data-stores, social apps, banking systems, videos and streaming services, news websites, etc.
The common motivation behind caching is to store data closer to the application than its source.
For example, DRAM memory is faster and closer in the memory hierarchy than secondary storage, local storage is closer than accessing data over the Internet, etc.
This way, the requested information can be served quickly whenever the requested data is already in the cache, also known as a \emph{cache hit}.
In the case of distributed/networked systems, caching also saves network bandwidth and reduces the load on servers.
For instance, when a user opens a cached news website, there is no need to load it from the~webserver.
\fi

Since caches are limited in size, a \emph{cache management mechanism} is required to decide which items should be kept in the cache. 
Similarly, when there is not enough space for all items, the management scheme decides which items should get evicted (known as \emph{cache victims}).
A plethora of cache management policies has been devised, including e.g., LRU~\cite{LRU}, LFU~\cite{LFU}, ARC~\cite{megiddo2003arc}, LIRS~\cite{jiang2002lirs}, FRD~\cite{park2017frd}, Hyperbolic~\cite{hyperboliccache}, and W-TinyLFU~\cite{einziger2017tinylfu}, to name a few.
Largely speaking, these schemes treat the cache as a \emph{fully associative} structure.
That is, for every item that is inserted into the cache, the management policy can potentially select any other item as its~victim.

In contrast, hardware managed caches 
employ a limited associativity design.
That is, the CPU cache is divided into multiple sets, each of which contains $k$ locations, forming a \emph{$k$-way associative cache}.
This is to ensure bounded time look-ups and to reduce the hardware circuitry complexity and cost required to support full associativity.
\ifdefined\MWVERSION
\else
Recently, it was shown that even for software caches it makes sense to employ a limited associativity design~\cite{kway}. This is because limited associativity enables trivial parallelism, reduces contention on data structures, increases memory density, and enables very simple $O(1)$ inserts and look-ups.
\fi

In-network caching promotes storing cached data in network switches and routers.
Since the data flows through these devices in any case, it makes sense to cache it there.
In fact, in-network caching is considered one of the enabling technologies for 5G and 6G's promised performance boost as well as edge computing.

Programmable switches are gaining momentum, with offers from large vendors such as Intel~\cite{tofino}, Nvidia/Mellanox~\cite{spectrum}, and Broadcom~\cite{broadcom}.
Instead of just forwarding packets, the switch can be programmed to manipulate and monitor the data flowing through it.
Yet, for speed and energy efficiency, this programming model is restricted compared to general-purpose computing~\cite{bosshart2014p4,P4spec}.

The pioneering \NetCache~\cite{netcache} implements a key-value distributed cache inside programmable switches.
\NetCache{} processes queries for hot items and balances the load across the storage nodes, while each caching switch is treated as a fully associative structure, implementing an ad-hoc cache management policy.
We claim that limited associativity designs are better fitted to programmable switches’ goal of handling packets in a short bounded amount of time and their restricted programming model.
Limited associativity also simplifies the realization of existing well studied cache management approaches inside the programmable switch.

\paragraph{Contributions:} We study the utility of limited associativity caching in programmable switches.
In particular, we present \PKache, a novel generic \PFOUR{} caching framework that adheres to $k$-way associativity cache design, where $k$ is a controllable compile time parameter.
\PKache{} supports both single region caching as well as multi-region caching, and can be instantiated with diverse specific cache management schemes in both cases.
In particular, we have implemented LRU~\cite{LRU}, LFU~\cite{LFU}, FIFO, and Hyperbolic~\cite{hyperboliccache} policies for the single region case, as well as the popular W-TinyLFU~\cite{einziger2017tinylfu} as a representative of multi-region schemes.
Many other cache management policies can be similarly implemented.
As we explain in this paper, $k$-way associativity helps overcome the limitations presented by the \PFOUR{} programming model.

We have compared the hit-ratios obtained by our \PKache{} system to those obtained by an unrestricted Python implementation of the respective schemes over several synthetic and real-world traces.
The results indicate that despite the restrictions of the \PFOUR{} programming model, \PKache{} yields very similar hit ratios to an unrestricted~implementation.

\ifdefined\MWVERSION
\else
\paragraph{Paper roadmap:} We provide more detailed background and survey related work in Section~\ref{Background and Related Work}. The \PFOUR{} Programming Language is briefly presented in Section~\ref{P4 Programming Language}. We present \PKache{} in Section~\ref{Design and Implementation} and evaluate the obtainable performance in Section~\ref{Experimental Results}. We conclude with a discussion in Section~\ref{Conclusions and Discussion}.
\fi

\begin{figure}[t]
  \centering
  \begin{subfigure}[b]{0.7\textwidth}
    \includegraphics[width=0.95\linewidth]{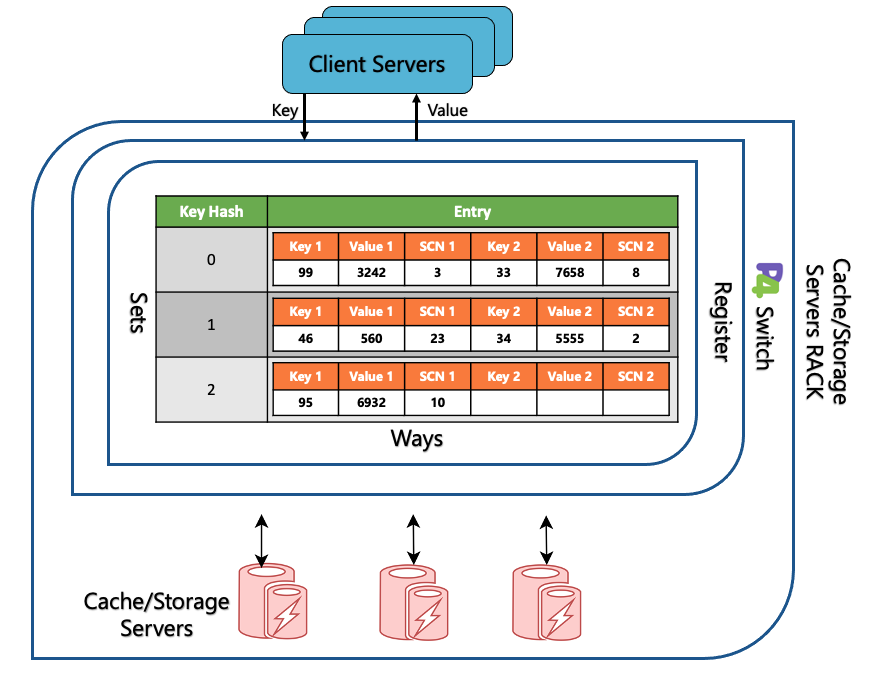}
  \end{subfigure}
  \caption{\PKache{} stores in a \PFOUR{} register $d$ sets, where each set contains $k$ ways (here $d=3$ and $k=2$). Each element includes a key, a value and SCN (Sequence Change Number) - the latter encodes any additional information needed for the cache management policy). }
  \label{fig:schema}
\end{figure}

\section{Background and Related Work} \label{Background and Related Work}
\subsection{Programmable Switches}
Programmable switches are capable of performing more sophisticated tasks than the ordinary packet forwarding action, based on a dynamically uploaded program.
That is, their data plane functionality can be defined by the given~program.

One of the most popular programming languages for such data planes today is \PFOUR.
\PFOUR{} is a domain-specific language and it is also a target specific language, so in every target, the spec of the language can be a little bit different, depending on the manufacturer of the switch.  \PFOUR{} Supports conditions, match-action tables, state-full objects, and more.  It does not support very common control methods like loops, recursion, etc.  The match-action table is a list of conditions. When one of them is fulfilled, a predefined action is executed.  This action performs some business logic on the incoming~packet. 

\subsection{Cache Management Policies}
As mentioned in the Introduction, there are numerous cache management policies.
Here we briefly describe the five
policies whose $k$-way variants we implement in our work.

\paragraph*{Least Recently Used (LRU)~\cite{LRU}}
LRU is arguably the most widely employed policy.
\ifdefined\MWVERSION
When the cache is full, the algorithm always evicts the least recently used item.
\else
LRU is based on the underlying principle of \emph{time locality}, suggesting that the probability of accessing a given item is related to the time that has passed since it was last accessed.
Hence, when the cache is full, the algorithm always evicts the least recently used item.
\fi
The simplest way to implement LRU is through a priority queue where each time an item is being accessed, it is moved to the head of the queue.
Yet, such an implementation creates high contention on the queue's head.
It also requires updating the cache meta-data on each access (both hits and misses).

\emph{Sampled LRU}~\cite{redis} is an approximate alternative, in which (only) the timestamp of each item's last access is stored.
To find the cache victim, a sample of $k$ random items is selected, and the least recently accessed among them becomes the victim.
Alas, this still requires updating the meta-data on each access, invoking the PRNG $k$ times, and accessing $k$ random memory~locations.

Another popular approximation of LRU is \emph{Clock}~\cite{clock}, in which the system treats the cache as a logical ring and lazily resets the access time of one item on each cache access using an analogy of a clock's moving hand.
The victim is the first item whose last access time is zero.
Here again, the meta-data needs to be updated on each cache access, and the worst-case eviction time is $O(C)$ for a $C$-sized cache.
Also, being approximate, both sampled LRU and Clock exhibit slightly worse hit-ratios than LRU.

\paragraph*{Least Frequently Used (LFU)~\cite{LFU}}
LFU is based on the assumption that the probability of an item being accessed is proportional to its popularity, or in other words, to the frequency by which it was accessed until now.
To that end, LFU maintains a frequency counter for each cached item. 
When the item is accessed, the counter is incremented by $1$.
When the cache is full and a new item needs to get in, the victim is the cached item with the minimum~frequency.

Clearly, for LFU the size of the counters can be significant over time.
There are a few mechanisms to decrease the values of the counters once in a while as well as to age the counters to accommodate for recency effects.
\ifdefined\MWVERSION
\else
A naive implementation of LFU uses a heap data structure, whose complexity is $O(\log C)$.
More recently, it was shown how to implement a heap suitable for LFU in $O(1)$~\cite{WCSS}.
\fi

\paragraph{FIFO~\cite{Belady}}
With the FIFO policy, the cache behaves in the same manner as a FIFO (First In, First Out) queue. 
That is, the cache evicts the elements in the order they were added, regardless of how often or how recently they were accessed before.
The benefits of this policy include its simplicity and the fact that an item's meta-data is never updated after its initial insertion to the cache.
Alas, its performance is often worse than other policies and it may suffer from the Belady anomaly in some workloads~\cite{Belady}.
Still, a recent work suggests that for modern cloud storage workloads, its performance is comparable to LRU~\cite{FIFOLRU}, and sometimes even better.

\paragraph*{Hyperbolic Cache~\cite{hyperboliccache}}
The recent Hyperbolic cache policy combines two metrics: recency and frequency.
When a new item enters the cache, the insert time is logged and a new counter is initiated for the newly cached item with a value of $1$.
The algorithm increments this counter by $1$ every time the item is requested. 
When the cache is full, the eviction mechanism samples a few items and calculates their relative priority.
The victim is the item whose priority is the lowest among the sampled items.
Formally, denote the insert time of item $i$ as $t_i$ and the request count of $i$ as $n_i$.
The priority of item $i$ at time $\mathit{now}$ is calculated as $p(i) = \frac{{n_i}}{\mathit{now} - t_i}$.

\begin{figure}[t]
  \centering
    \includegraphics[width=0.7\linewidth]{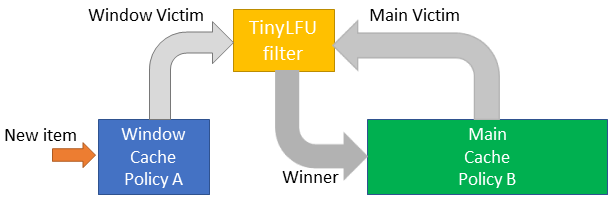}
  \caption{\TinyLFU{} Schematic Overview}
  \label{fig:tinylfu}
\end{figure}

\paragraph*{\TinyLFU{}~\cite{einziger2017tinylfu}}
\TinyLFU{} maintains two cache regions, a \WindowCache{} and a \MainCache{} as well as an approximate frequency-based admission filter called TinyLFU.
With \TinyLFU{}, new items are first inserted into the \WindowCache{}.
Victims of the \WindowCache{} are compared by the TinyLFU filter against the would-be victim of the \MainCache{} in terms of their approximate frequency.
The winner gets to be in the \MainCache{}, while the loser is completely removed from the cache (although its approximate statistics is still tracked by the TinyLFU filter). See illustration in Figure~\ref{fig:tinylfu}.

The TinyLFU filter maintains an approximate representation of the access frequency of a large sample of recently accessed items. 
\TinyLFU{} is compact and lightweight as it is based on Count-Min sketch~\cite{cormode2005improved}. 
It includes an aging mechanism and it caps the maximal frequency counts, to expedite the aging of items. 
\TinyLFU{} is the management policy employed by the Caffeine~\cite{CaffeineProject} and Ristretto~\cite{RistrettoProject} caching libraries, and many other projects and products.

\subsection{Caching in the Data Plane}

\paragraph*{\NetCache{}~\cite{netcache}}
\NetCache{} is a rack-scale key-value store design that supports billions of QPS with bounded latency.
\NetCache{} includes an internal load balancer to help it override spikes in the number of requests. 
Furthermore, \NetCache{} guarantees cache coherence with only a minor overhead.
\NetCache{} is implemented in \PFOUR{}, and its performance exemplifies the benefits to distributed systems enabled by high-speed programmable switches. 
We note that the evaluation part of~\cite{netcache} only focused on throughput and latency and did not include any hit ratio measurements.

The cache management mechanism employed by \NetCache{} is an ad-hoc policy hard-coded into the solution.
In contrast, \PKache{} is a generic caching architecture for programmable switches.
Also, \PKache{} demonstrates the flexibility and simplicity that $k$-way associativity brings to in-network caches implemented inside programmable switches.

\ifdefined\MWVERSION
\else
\paragraph*{Limited Associativity Software Cache~\cite{kway}}
%
The work in~\cite{kway} demonstrated that limited associativity is a promising method for software caches. 
It is easier and simpler to implement a limited associativity cache than a fully associative one.
Using limited associativity reduces memory overheads and increases parallelism compared to fully associative memory.
The resulting cache also imposes lower CPU overheads than fully associative designs and sampled based approaches.
The major penalty of this method is a slight reduction in hit ratio compared to full associativity in certain workloads. 
\fi

\paragraph*{KV Direct~\cite{li2017kv}}
KV Direct offers a smart NIC library for accessing an in-memory key value store using RDMA.
Hence, KV Direct~\cite{li2017kv} expedites accesses to nearby machines, connected through a very fast network, while our work provides a general in-network caching solution for key-value stores, even remote ones.
%

\section{P4 Programming Language} \label{P4 Programming Language}
\PFOUR{} is a domain-specific language describing how a PISA~\cite{pisa} (Protocol Independent Switch Architecture) realization should process its packets. 
Also, it is a target-specific language, so in every target, the spec of the language can be a little bit different, depending on the manufacturer of the switch. 
The main motivation behind \PFOUR{} is to enable expressing how packets should be processed by the data plane of a programmable forwarding element such as a hardware or software switch, network interface card, router, or network appliance~\cite{bosshart2014p4,P4spec}.

Unlike most programming languages, \PFOUR{} does not support common features such as loops, recursive calls, memory allocation, or even the use of division and floating points. 
This is meant to ensure that the amount of computation per packet can be bounded at compilation time and all data structures' sizes must be fixed at compile time.
\ifdefined\MWVERSION
\else
On the other hand, \PFOUR{} supports some special constructs like tables, parsers, deparsers and more, which we survey below.

\subsection{\PFOUR{} Constructs}
\begin{description}
\item[Parsers:]
Parser is a section that describes how to identify the packets' header, how to extract them, and the permitted sequences we can perform on them. 
It uses a state machine to extract from the packet its header and has three predefined states: Start, Accept and Reject.

\item[Deparsers:]
Deparser describes how the output packet would look like. It emits the headers into a well-formed packet that will be returned from the~switch.

\item[Actions:]
Actions are code sections that can read and write the data being processed and are the main section in which the control plane can influence the data plane. They are like procedures in other languages.
The body consists of sequences of statements. 
Note that some targets, like BMv2~\cite{bmv2}, impose restrictions on the actions' body, e.g., conditional execution in actions is not supported on BMv2.

\item[Tables:]
Tables are a special structure that describes a match-action unit. 
Processing a packet using a match-action table involves the following steps: key construction, key lookup in a lookup table (``match'' step) and an action execution (``action'' step)~\cite{P4spec}. 
A \PFOUR{} table consists of the following: 

\begin{description}
\item[Key:] A key is a property in the form of (e~:~m) when e is an expression that describes the data to be matched in the table and m is a $match\_kind$ constant that describes the algorithm used to perform the lookup. 
There are three kinds of match types. 
One of them is called $ternary$.
$Ternary$ has the following meaning: On each entry in the table, a mask is provided. 
Then, the field value and the mask are bit-wise ANDed before a comparison is made. 
The field value and the table entry need to agree on the bits set in the entry's mask. 
Agreement means that all the ``regular'' bits are the same between the result and the entry.
\ifdefined\MWVERSION
\else
We can create masks in \PFOUR{} using the infix operator \&{}\&{}\&{}.
It takes two arguments of type bit\textlangle W\textrangle\ and creates a value of type set bit\textlangle\textlangle W\textrangle\textrangle. 
The right value is used as a mask, where each bit set to 0 in the mask is a ``don't care'' bit. 
Mathematically, we define \&{}\&{}\&{} in the following way:\\
\small
$a\ \&{}\&{}\&{}\ b\ =\ \{\ c\ \mathit{of\ type\ bit}<W>\ \mathit{where}\ a\ \&{}\ b\ =\ c\ \&{}\ b\ \}$.
\normalsize
\fi
\item[Actions:] All possible actions that may appear within the associated lookup table or in the default action.
\item[Default action:] An action that is invoked automatically by the match-action unit whenever there is no match for the supplied key in the entries section.
\item[Entries:] An entry is a list of table properties in the form of ($v$~:~$a$) where $v$ is a value and $a$ is an action. 
When the value is matched by the $match\_kind$ algorithm, the desired action is performed. 
Entries in a table are matched in the program order, stopping at the first matching entry.
\end{description}

\item[Stateful Objetcs:]
Most \PFOUR{} constructs are stateless: 
Given an input packet, they produce a result that depends only on the packet itself. 
However, some stateful constructs may retain information across packets. 
One type of stateful object is called $extern$. 
These are objects that can be read and written by the control plane and data plane. 
As we mentioned before, all stateful elements must be explicitly allocated at compilation time and have a fixed size.

One of the $extern$ objects is called a Register. 
Registers are stateful and behave like arrays (index and value).
We can reference an item in a register by its index using the $register\_read$ and $register\_write$ primitives.
\end{description}
\fi

\subsection{Challenges with \PFOUR{}}
Unlike other programming languages, \PFOUR{} does not support loops, recursive calls, floating-point operations and more. 
Without loops, iterating over registers to determine if an element is inside the cache or not is hard, not to mention the eviction process in which we need to check each element to determine which element should be evicted according to the cache policy. 
Any code that needs to iterate over a register's values is impossible, and we have to find~workarounds.

The lack of floating point operations complicates certain algorithms and implementing probabilistic structures. 
In Hyperbolic cache~\cite{hyperboliccache}, for example, we have to find an approximate way to calculate the priority of each item to determine the preferred element for~eviction.

\subsection{\PFOUR{} and Limited Associativity Caching}
We argue that \PFOUR's design and restrictions as surveyed above encourage us to favor a limited associativity design over a fully associative one.
The lack of loops and iterations makes it very inefficient to work with fully associative caches in terms of run time and the number of operations.
Further, the lack of a loop construct means that an inlined code needs to be generated instead, which would result in significant code bloat.
In principle, the use of TCAM enables us to determine whether an item is in the cache or not in a single comparison.
However, as we detail below, there is a limit on the size of a TCAM mask and therefore on the number of items that can be matched in a single TCAM comparison.


In contrast, a limited associativity cache overcomes, or at least ameliorate, these limitations.
With such a design, we are able to deploy large caches while minimizing the impact on the runtime or the number of operations performed per packet.
For example, when we organize the cache as a $k$-way, the TCAM comparison only needs to match the $k$ items inside the set to which the item is mapped.
This is independent of the number of sets in the cache.
Hence, we can increase the cache size by increasing the number of sets while keeping $k$ relatively small.
The exact details are explained~below.

\section{Design and Implementation} \label{Design and Implementation}
\subsection{Overview}
\PKache{} is implemented in \PFOUR{} spec 16 \cite{P4spec, bosshart2014p4} and tested on BMv2~\cite{bmv2}. 
We implemented \PKache{} to work with both a single cache region and multiple cache regions designs. 
Currently, \PKache{} supports LFU~\cite{LFU}, LRU~\cite{LRU}, Hyperbolic~\cite{hyperboliccache} and FIFO as cache management policies for each cache region as well as a TinyLFU filter in case of multiple cache regions, thereby also supporting the \TinyLFU{} scheme~\cite{einziger2017tinylfu}.
The \PFOUR{} code is generated by a Python script~\cite{PKache-code} using \emph{Jinja2}~\cite{jinja} and depends on the following parameters (i.e., these values cannot be changed during run-time):
\begin{itemize}
    \item $k_m$ - Number of ways in the \MainCache{}.
    \item $d_m$ - Number of sets in the \MainCache{}.
    \item $k_w$ - Number of ways in the \WindowCache{} (in case of multi cache region).
    \item $d_w$ - Number of sets in the \WindowCache{} (in case of multi cache region).
\end{itemize}

For any incoming packet with a key request, our algorithm checks if the key exists in the cache.
In case the key exists, the value is fetched and returned to the client inside an outgoing packet, without reaching the storage servers at all.
Otherwise, the switch routes the packet to the storage servers and waits for a response. 
When the response is transmitted to the client through the switch, our algorithm updates the cache depending on the policy and routes the outgoing packet to the client. 
In a multi-region cache, the process is slightly different, as detailed in Section~\ref{multi} below. 
We now cover the main components of our solution:

\begin{description}
\item [In-Switch Storage:] Our memory-store consists of one \PFOUR{} register for each cache region. 
A \PFOUR{} register is a stateful predefined fixed-sized object with key-value store capabilities.

\PKache{} stores sets of cached items inside a register. 
The keys for the register's entries are numbered from $0$ to the number of sets ($d_m$ or $d_w$) minus 1.
Items are mapped to sets by hashing their keys to the range of sets;
we refer to the set index as the Key Hash.
An illustration of a \PKache{} register is shown in Figure~\ref{fig:schema}.

The value of each entry in the register is the cache set itself, which is a bit array containing all items key-value pairs of the set.
Those pairs are stored in the memory space. 
Further, we allocate some extra auxiliary space for each key-value pair as a dedicated area for the cache policy metrics. We refer to the area reserved for the policy metrics as \emph{Sequence Change Number} (SCN).

The algorithm parses the data with a bit-slice operator, so the exact key-value-metric can be easily represented as attributes.
Also, we use additional stateful registers for our implementation. 
These include one register for each cache region that is built like the previous one but contains just the keys (and not the whole element).
This register helps us determine whether the key is inside the cache or not. 
Another register represents the potential candidate for eviction and some specific policy implementation registers.

\item [Pipeline Model:] Programmable switches offer the pipelined PISA execution model and \PFOUR{} exposes the pipeline architecture to the programmer. 
Each \PFOUR{} code can be broken into stages, and different stages can be run in parallel to improve performance.
As a rule of thumb, the number of stages should be limited to a few dozens so the switch latency is not detrimentally affected.

\item[Ternary Operations:] For any new item, we find the relative set that corresponds to the desired key. 
Then, our algorithm uses a table with a ternary match\_kind identifier (TCAM) that indicates whether the requested key is inside the cache or not. 
This is performed by taking the value in the keys register for the corresponding set and using a bit-wise XOR operation with the requested key concatenated $k$ times.
Then we are using a ternary mask to check whether one of the values in the TCAM entries is equal to zero. 
If it is indeed the case, i.e., the key is found in the cache set, then we know that the item is in the cache; otherwise, it is not.

Thus, with just one table comparison and without the need to iterate over the cache elements, we can discover if the requested key is in the cache.
In addition, we know exactly the key's position inside the cache (the TCAM entry whose value was equal to zero), so we are able to get the item directly from the cache when it exists there.
This is made possible by the limited associativity approach.
Note that here we assume that the requested key is not equal to zero.

\item[Reduce Processing:] For inserting an element to the cache, we are using the reduction (or folding) pattern.
In our case, we reduce the cache set items to a single cumulative value (the element or the victim). 
Specifically, we apply our comparison function to the first element's key (among the $k$) in the respective set and the requested key to generate a partial result. 
We then use our partial result together with the second item in the set to generate another partial result.
This is repeated until the set is exhausted and then we get a single cumulative value (for example, the candidate for eviction).
Obviously, here the fact that $k$, the number of items per set, is relatively small is important to ensure fast packet processing.

\item [Eviction Process:] 
In case the key exists, the algorithm fetches the value from the cache and returns it in an outgoing packet to the client. 
Otherwise, the algorithm fetches the key from the storage server and stores the key-value pair in the first empty way of the respective set in the cache.
In case the cache set is full, based on the chosen cache management policy, the candidate for eviction is chosen by comparing the metrics values of all elements in the set using the reduction technique we described above.
Note that in the case of a multi-region cache, the eviction process is more complicated, so we describe it later. \label{eviction}
\end{description}

\subsection{Single Cache Region}
\subsubsection{FIFO Single \PKache{}}
Algorithm~\ref{alg:fifo} lists the code for \PKache{}'s FIFO policy implementation when inserting an element. 
Initially, the algorithm inserts the element to the first place in the respective set (lines~\ref{alg:fifo:insertend}-\ref{alg:fifo:insertstart}). 
By doing so, we take the first element that was there before and mark it as the candidate (line~\ref{alg:fifo:swap}). 
We then apply the reduction process in which we insert the candidate to the cache and mark the next element in the set as the new candidate (lines~\ref{alg:fifo:procstart}-\ref{alg:fifo:procend},\ref{alg:fifo:reducestart}-\ref{alg:fifo:reduceend}).
At the end of the process, the last element (the first element that was inserted and is still in the cache) leaves the cache.

Note that FIFO serves as a relatively easy exercise for implementing cache management policies.
Still, recent evidence suggests that on many modern workloads, FIFO can obtain comparable (sometimes even better) hit ratios to LRU~\cite{FIFOLRU}.

\begin{algorithm}
\caption{Insert element to a $4$-way cache with FIFO policy (automatically generated for $k_m=4$)}
\label{alg:fifo}
\footnotesize
\begin{algorithmic}[1]
\State $Initialization:$
\ifdefined\MWVERSION
\State\tab $cache \gets Register(d_m)$; $keys\_cache \gets Register(d_m)$
\else
\State\tab $cache \gets Register(d_m)$
\State\tab $keys\_cache \gets Register(d_m)$
\fi
\newline
\Procedure{get\_candidate}{$cache\_set$,$keys\_set$,$index$,$candidate$}
    \State $temp \gets cache\_set[index]$ \label{alg:fifo:procstart}
    \State $cache\_set[index] \gets candidate$
    \State $keys\_set[index] \gets candidate[KEY\_LOC]$
    \State $candidate \gets temp$
    \State\Return $candidate$ \label{alg:fifo:procend}
\EndProcedure
\newline
\Procedure{insert\_to\_cache}{$key$}
    \State $h \gets key \%{}d_m$ 
    \State $cache\_set \gets cache[h]$
    \State $keys\_set \gets keys\_cache[h]$
    \State $candidate \gets cache\_set[0]$ \label{alg:fifo:insertstart} \label{alg:fifo:swap}
    \State $cache\_set[0][KEY\_LOC] \gets pkt.key $
    \State $cache\_set[0][VAL\_LOC] \gets fetch\_value\_from\_storage() $
    \State $keys\_set[0] \gets pkt.key$ \label{alg:fifo:insertend}
    \State $candidate \gets$ \Call{get\_candidate}{cache\_set, keys\_set, 1, candidate} \label{alg:fifo:reducestart}
    \State $candidate \gets$ \Call{get\_candidate}{cache\_set, keys\_set, 2, candidate}
    \State\Call{get\_candidate}{cache\_set, keys\_set, 3, candidate} \label{alg:fifo:reduceend}
    \State\Return $cache\_set[0]$
\EndProcedure
\end{algorithmic}
\end{algorithm}

\subsubsection{LRU Single \PKache{}}
\ifdefined\MWVERSION
\else
LRU needs to keep track of items' most recent access to evacuate the least recently used item.
\fi
Algorithm~\ref{alg:lru} lists the code for our \PKache{}'s LRU policy implementation.
Recall that this code is automatically generated for any value of $k_m$; in this example, $k_m=3$.
The implementation uses a timestamp to record when each item was recently accessed. 
Our algorithm maintains this timestamp in the Sequence Change Number (\SCN{}) field. 
The \SCN{} is a number that is incremented every time there is an update. 
Thus, every change has a unique number that is distinguishable from other changes. 
By using \SCN{} as the last access indicator, it is easy to decide which item has not been accessed for the longest period.
Given that each set has only a small number of items, this can be done efficiently (a benefit of limited associativity).

First, the algorithm checks if the key exists in the cache (line~\ref{alg:lru:check}) using the \KeysSet{} and the ternary match table. 
As mentioned, we concatenate the key \emph{cache\_size} times to get a bit string, and then xor the result with the \KeysSet{} represented as a bit string. 
If any index returns zero, it means that the element is inside the cache. 

If the key is inside the cache, we are fetching the value immediately from the relevant index in the \CacheSet{}.
At this point its \SCN{} is updated to the value defined in line~\ref{alg:lru:scn} and the cached value is returned to the client (lines~\ref{alg:lru:startget}-\ref{alg:lru:endget}).

If the key does not exist in the cache at all, we insert the key as the first element in the \CacheSet{} (and update the \KeysSet{} accordingly).
We also define the previous first element as our victim for eviction (lines~\ref{alg:lru:firstvictimstart}-\ref{alg:lru:firstvictimend}).
The algorithm fetches the value from the server cache or storage. 
We define an abstract method for this step in line~\ref{alg:lru:fetch}, but it can be easily changed.

After that, we are iterating over the \CacheSet{}.
At each step, we compare the candidate's \SCN{} to the current element's \SCN{} in the set. 
If the \SCN{} is lower (which means the candidate is not the least recently used element), we swap the current element and the candidate in the set (lines~\ref{alg:lru:startinsert}-\ref{alg:lru:endinsert}).

\begin{algorithm}[t]
\caption{A LRU $3$-way cache (automatically generated for $k_m=3$)}
\label{alg:lru}
\footnotesize
\begin{algorithmic}[1]
\State $Initialization:$
\ifdefined\MWVERSION
\State\tab $cache \gets Register(d_m)$; $SCN \gets 0$
\else
\State\tab $cache \gets Register(d_m)$
\State\tab $SCN \gets 0$
\fi
\newline

\Procedure{get\_element}{$pkt.key$, $cache\_set$, $index$, $SCN$, $element$}
    \If{ $cache\_set[index][KEY\_LOC] == pkt.key$ }
        \State $cache\_set[index][SCN\_LOC] \gets SCN$
        \State $element \gets cache\_set[index]$ 
    \EndIf
    \State\Return $element$ 
\EndProcedure
\newline

\Procedure{insert\_element}{$cache\_set$, $keys\_set$, $index$, $candidate$}
    \If{ $cache\_set[index][SCN\_LOC] < candidate[SCN\_LOC]$ }
        \State $temp \gets cache\_set[index]$
        \State $cache\_set[index][KEY\_LOC] \gets candidate[KEY\_LOC] $
        \State $cache\_set[index][VAL\_LOC] \gets candidate[VAL\_LOC] $
        \State $cache\_set[index][SCN\_LOC] \gets candidate[SCN\_LOC] $
        \State $keys\_set[index] \gets candidate[KEY\_LOC] $
        \State $candidate \gets temp$
    \EndIf
    \State\Return $candidate$
\EndProcedure
\newline

\Procedure{fetch}{$pkt$}
    \State $h \gets pkt.key \%{}d_m$ 
    \State $cache\_set \gets cache[h]$
    \State $keys\_set \gets keys\_cache[h]$
    \State $SCN \gets SCN + 1$  \label{alg:lru:scn}
    \If{ $index \leftarrow keys\_set.contains(pkt.key)$ } \label{alg:lru:check}
        \State $element \gets None$  \label{alg:lru:startget}
        \State $element \gets$ \Call{get\_element}{pkt.key, cache\_set, index, SCN, element}
        \State\Return $element$ \label{alg:lru:endget} 
    \Else
        \State $candidate \gets cache\_set[0]$ \label{alg:lru:startinsert} \label{alg:lru:firstvictimstart}
        \State $cache\_set[0][KEY\_LOC] \gets pkt.key $
        \State $cache\_set[0][VAL\_LOC] \gets fetch\_value\_from\_storage() $ \label{alg:lru:fetch} 
        \State $cache\_set[0][SCN\_LOC] \gets SCN $
        \State $keys\_set[0] \gets pkt.key$ \label{alg:lru:firstvictimend}
        \State $candidate \gets \Call{insert\_element}{cache\_set, keys\_set, 1, candidate}$
        \State\Call{insert\_element}{cache\_set, keys\_set, 2, candidate}
        \State\Return $cache\_set[0]$  \label{alg:lru:endinsert}
    \EndIf
\EndProcedure
\end{algorithmic}
\end{algorithm}

\subsubsection{LFU Single \PKache{}}
LFU maintains the frequency access of each key. 
Our LFU implementation is very similar to LRU, except for the \SCN{} handling. 
Here, \SCN{} is incremented each time the element is accessed (frequency count). 
Since we increment it on each access, we need to add some aging to this mechanism (because we measure the last frequencies and not the total count).
Therefore, we decrement the element's \SCN{} when other elements are accessed.

\subsubsection{Hyperbolic Single \PKache{}}
Recall that in hyperbolic caching, an item's priority is set to the number of times the item was accessed since inserted into the cache divided by the duration of time the item remains in the cache.
The item whose priority is minimal is deemed the cache~victim.

Alas, \PFOUR{} targets usually support only simple arithmetic operations, such as addition and subtraction, but not multiplication and division.
This makes implementing Hyperbolic caching more challenging.
Even though such operations exist in the \PFOUR{} spec~\cite{P4spec}, popular targets like Intel's Tofino~\cite{tofino} and Mellanox' Spectrum~\cite{spectrum} do not support them.
The main reason is the latency and energy costs of these operations.
A multiplication runs many instructions on the CPU compared to addition and subtraction, and in real-time hardware like a switch, every instruction counts.
Hence, to implement Hyperbolic caching in \PKache{}, we use two different approaches:
\begin{description}
    \item[Semi-Division:] Obtaining an item's priority in Hyperbolic caching requires division.
    As mentioned above, division is not supported on \PFOUR{} targets.
    Following~\cite{pint}, we rewrite the division $\frac{X}{Y}$ as the following equation: 
    \small
    $$\frac{X}{Y} = 2^{\log_2(x) - \log_2(y)}.$$
    \normalsize
    In our case, the entire result is not significant. 
    Rather, the goal is to compare the results and free the cache slot whose value is minimal. 
    Thus, in \PKache{} we only care about the $\log_2(x) - \log_2(y)$ part of the equation.
    \PFOUR{} spec does not provide a $\log$ operation either, but it provides a cache and lookup tables.
    Our solution involves an extra auxiliary space that is initialized during deployment and includes the $\log_2$ values of a specific range of numbers.
    This means that we add a small error since we cannot represent all values.
   
    In summary, we replace the division phase of the counter by the insert time with the subtraction of their respective $\log$ values.
    The log values are stored in a register that is initiated at deployment. 
    This is useful and efficient since the register is used for both sides of the fraction.
    In case the value is larger than the maximal integer number in the register, the algorithm uses the maximum stored value.
    To limit the impact of large numbers, every time we reach the maximum, all values are divided by $2$ using a shift right operation, which is supported by the \PFOUR{} spec and targets.

    \item[Integer Factor:] A $\log_2$ operation returns a floating-point for most numbers. 
    Floating points are not supported in \PFOUR, therefore most of the results we aim to store are indistinguishable in their integer form.
    Instead, each floating-point is converted into an integer number while a piece of the fractional part is represented in an integer form with a pre-defined accuracy error.
    For instance, storing a number with an accuracy of 2 digits after the decimal point transforms the floating-point $123.45678$ into the integer $12345$.
\end{description} 
Algorithm~\ref{alg:hyperbolic} depicts \PKache's Hyperbolic caching scheme's code that is generated for a given $k_m$ ($k_m=3$ in this example).
In Algorithm~\ref{alg:hyperbolic}, the first part of fetching a cached value is similar to the previous algorithms we have seen (lines~\ref{alg:hyperbolic:startget}-\ref{alg:hyperbolic:endget}). In case the key exists, we increment its frequency by~1.

Otherwise, we insert the key as the first element like previous algorithms, then set its frequency to 1 and its insertion time to $SCN$. 
We also define the previous first element as our victim for eviction (lines~\ref{alg:hyperbolic:firstvictimstart}-\ref{alg:hyperbolic:firstvictimend}).

Next, we iterate over the \CacheSet{}. 
At each step, based on the lifetime period of the elements and their frequency, the algorithm fetches the estimated $\log$ values for the candidate and the current element (floating points are converted to integers using the \emph{integer factor} approach). The initialization of the $log\_vals$ register is done in line~\ref{alg:hyperbolic:init}. The packet that we send is in the form of \emph{(type=LOG\_INITIALIZATION, key=i, value=}$\lfloor\log_2(x)\cdot \mathit{integer factor}\rfloor$\emph{)}. 
The priority is defined by subtracting two $\log$ values (lines~\ref{alg:hyperbolic:startcalcprior}-\ref{alg:hyperbolic:endcalcprior}).
Then, the algorithm compares both the candidate's and the current element's priority values and if the candidate's priority is higher, we swap the current element and the candidate in the set (lines~\ref{alg:hyperbolic:startcompprior}-\ref{alg:hyperbolic:endcompprior}).

\begin{algorithm}
\caption{A Hyperbolic $3$-way cache (automatically generated for $k_m=3$)}
\label{alg:hyperbolic}
\footnotesize
\begin{algorithmic}[1]
\State $Initialization:$
\ifdefined\MWVERSION
\State\tab $cache = Register(d_m)$; $packet\_counter = 0$;
\else
\State\tab $cache = Register(d_m)$
\State\tab $packet\_counter = 0$
\fi
\State\tab $log\_vals = Register(MAX\_SCN)$
\newline

\Procedure{get\_element}{$pkt.key$, $cache\_set$, $index$, $element$}
    \If{ $cache\_set[index][KEY\_LOC] == pkt.key$ }
        \State $cache\_set[index][FREQ\_LOC] \gets cache\_set[index][FREQ\_LOC] + 1$
        \State $element \gets cache\_set[index]$ 
    \EndIf
    \State\Return $element$ 
\EndProcedure
\newline

\Procedure{insert\_element}{$cache\_set$, $keys\_set$, $index$, $candidate$, $scn$}
    \State $element\_freq \leftarrow  cache\_set[index][FREQ\_LOC]$ \label{alg:hyperbolic:startcalcprior}
    \State $element\_lifetime \leftarrow scn -  cache\_set[index][INSERTIME\_LOC]$
    \State $candidate\_freq \leftarrow  candidate[FREQ\_LOC]$
    \State $candidate\_lifetime \leftarrow scn -  candidate[INSERTIME\_LOC]$
    
    \State $p_{element} \leftarrow log\_vals[element\_freq] - log\_vals[element\_lifetime]$
    \State $p_{candidate} \leftarrow log\_vals[candidate\_freq] - log\_vals[candidate\_lifetime]$ \label{alg:hyperbolic:endcalcprior}
    \If{ $p_{element} < p_{candidate}$ }
        \State $temp \gets cache\_set[index]$ \label{alg:hyperbolic:startcompprior}
        \State $cache\_set[index][KEY\_LOC] \gets candidate[KEY\_LOC] $ 
        \State $cache\_set[index][VAL\_LOC] \gets candidate[VAL\_LOC] $
        \State $cache\_set[index][FREQ\_LOC] \gets candidate[FREQ\_LOC] $
        \State $cache\_set[index][INSERTIME\_LOC] \gets candidate[INSERTIME\_LOCC] $
        \State $keys\_set[index] \gets candidate[KEY\_LOC] $ \label{alg:hyperbolic:endcompprior}
        \State $candidate \gets temp$
    \EndIf
    \State\Return $candidate$
\EndProcedure
\newline

\Procedure{fetch}{$pkt$}
    \If{$pkt.type is LOG\_INITIALIZATION$} 
        \State $log\_vals[pkt.key] = pkt.value$ \label{alg:hyperbolic:init}
        \State\Return
    \EndIf
    \State $h \leftarrow pkt.key \%{}d_m$ 
    \State $cache\_set \leftarrow cache[h]$
    \State $keys\_set \leftarrow keys\_cache[h]$
    \If{$index \leftarrow keys\_set.contains(pkt.key)$}
        \State $element \gets None$  \label{alg:hyperbolic:startget}
        \State $element \gets$ \Call{get\_element}{pkt.key, cache\_set, index, element}
        \State\Return $element$ \label{alg:hyperbolic:endget}
    \Else
        \State $candidate \gets cache\_set[0]$ 
        \State $cache\_set[0][KEY\_LOC] \gets pkt.key $         \label{alg:hyperbolic:firstvictimstart}
        \State $cache\_set[0][VAL\_LOC] \gets fetch\_value\_from\_storage() $  
        \State $cache\_set[0][FREQ\_LOC] \gets 1 $
        \State $cache\_set[0][INSERTIME\_LOC] \gets SCN $
        \State $keys\_set[0] \gets pkt.key$ \label{alg:hyperbolic:firstvictimend}
        \State $candidate \gets \Call{insert\_element}{cache\_set, keys\_set, 1, candidate, scn}$
        \State\Call{insert\_element}{cache\_set, keys\_set, 2, candidate, scn}
        \State\Return $cache\_set[0]$
    \EndIf
\EndProcedure
\end{algorithmic}
\end{algorithm}

\subsection{Multiple Cache Region} \label{multi}
\subsubsection{Differences from a Single Cache Region}
\PKache{} supports multiple cache regions, each with its own $d$ and $k$ values. 
As mentioned above, we refer to the first region as \WindowCache{} and to the second region as \MainCache{}.
The description below focuses on \TinyLFU{}, but with very few modifications it can be applied to other multi-region schemes like 2Q~\cite{2Q}.

In a multi-region cache, we have two ternary tables - one for each region. 
In case the key exists, the algorithm fetches the value from the corresponding cache region directly and returns it in an outgoing packet to the client.
Otherwise, the algorithm fetches the key from the storage server, and stores the key-value pair in the first empty set in the \WindowCache{}.

In case the \WindowCache{} set is full, based on the chosen \WindowCache{} policy, the algorithm chooses the candidate for eviction.
Then it takes the candidate and inserts it into the \MainCache{} if the \MainCache{} is not full. 
Otherwise, it will be inserted only if this element is better than one element in the \MainCache{} according to its policy. 

When we have a filter between the caches (for example, the TinyLFU filter), the element is inserted only if the candidate that was evicted from the \WindowCache{} is worse according to the filter than the candidate for eviction from the \MainCache{}. 
Note that since the \MainCache{} and \WindowCache{} may have different policies, our elements should maintain both \SCN{}s, so that we can compare elements from the \WindowCache{} and the \MainCache{} correctly (for example, LFU and LRU \SCN's behave quite~different).

\subsubsection{TinyLFU Filter}
\PKache{} supports a filter mechanism between the \WindowCache{} and the \MainCache{}. 
In particular, \TinyLFU{} employs the TinyLFU filter~\cite{einziger2017tinylfu}.
Given an eviction candidate from the \WindowCache{}, the filter decides based on the recent access history whether it is worth admitting the item into \MainCache{}. 

There are multiple ways to implement the TinyLFU filter. 
One efficient way is to use CM-Sketch~\cite{cormode2005improved}.
Yet, \PKache{} currently implements the filter using an explicit counting structure for all the elements with a de-amortized aging process~\cite{einziger2017tinylfu,CaffeineProject}.
The reason we choose to implement the filter with a counting structure instead of CM-Sketch is that our keys are quite small.
Further, since we are running on BMv2, we do not have a tight memory limitation. 
Yet, our implementation can be easily modified to use CM-Sketch, and \PKache{} is independent of the way the filter is~implemented.

As described in~\cite{einziger2017tinylfu}, we need to apply an aging mechanism to the filter to prevent once very popular items from polluting the cache after they stop being popular.
In the original \TinyLFU{} work, aging is performed by halving all counters once every $W$ accesses, where $W$ are some multiple of the cache size.
Alas, this means that once in a while, we have a very long aging operation, which would delay the packet whose treatment invokes this aging~process.

Hence, similarly to~\cite{CaffeineProject}, we de-amortize the aging mechanism to avoid long operations.
Specifically, once every $n << W$ accesses, we divide $\frac{{number\_of\_available\_keys \cdot n}}{{W}}$ counters by $2$, so that after $W$ turns, all items are divided by $2$. 
\subsubsection{Implementation}
Algorithm~\ref{alg:multi} lists the code for the multi-region cache with \TinyLFU. 
For clarity, we focus on the multi-cache regions and the filter and encapsulate the single cache regions implementation policies. 
The filter is shown in lines~\ref{alg:multi:filterstart}-\ref{alg:multi:filterend}. 
Note that \PKache's insert implementation starts with inserting the element to the first index (line~\ref{alg:multi:firstelement}), and then calculate the candidate from \MainCache{}. 
Therefore, we compare the candidate that we got from the \MainCache{} with the first element (the candidate from \WindowCache{}) to verify whether our insertion was correct.
If not, we will swap the elements.
In the latter case, it actually means that the element that is evicted is the element from the \WindowCache{} whereas the \MainCache{} remained intact.

\begin{algorithm}[t]
\caption{Multi Cache Implementation with \TinyLFU{}}
\label{alg:multi}
\footnotesize
\begin{algorithmic}[1]
\State $Initialization:$
\State\quad $main\_cache \gets Register(d_m)$
\State\quad $window\_cache \gets Register(d_w)$
\State\quad $counting\_structure \gets Register(number\_of\_available\_keys)$
\newline

\Procedure{fetch}{$pkt$}
    \State $h_{main} \gets pkt.key \%{}d_m$ 
    \State $h_{window} \gets pkt.key \%{}d_w$ 
    \State $main\_cache\_set \gets cache[h_{main}]$
    \State $window\_cache\_set \gets cache[h_{window}]$
    \State $counting\_structure[pkt.key] \gets counting\_structure[pkt.key] + 1$
    \If{ $main\_cache\_set.contains(pkt.key)$ }
        \State $element \gets get\_element\_from\_main\_cache(...)$
        \State\Return $element$
    \ElsIf{ $window\_cache\_set.contains(pkt.key)$ }
        \State $element \gets get\_element\_from\_window\_cache(...)$
        \State\Return $element$
    \Else
        \State $window\_candidate \gets window\_cache[0]$
        \State $window\_cache[0] \gets create\_new\_element(pkt.key, ...)$
        \State $window\_candidate \gets get\_candidate\_from\_window\_cache(...)$
        \State $main\_candidate \gets main\_cache[0]$
        \State $main\_cache[0] \gets window\_candidate$ \label{alg:multi:firstelement}
        \State $main\_candidate \gets get\_candidate\_from\_main\_cache(...)$
        \If{ $counting\_structure[candidate] > counting\_structure[main\_cache[0][KEY\_LOC]$ } \label{alg:multi:filterstart}
            \State $main\_cache[0] \gets candidate $
        \EndIf \label{alg:multi:filterend}
    \EndIf
\EndProcedure
\end{algorithmic}
\end{algorithm}

\subsection{Latency Analysis}
When a packet arrives we are performing the following:
\begin{itemize}
    \item One ternary match table per region to determine whether the key is in the cache or not. There is a read from the register to get the keys.
    \item In case of a cache hit, we require only one read from the register and one write to the register (read the whole set and then update the whole set after changing the SCN).
    Since the TCAM match action returns the element's position, we can access it directly.
    \item On a cache miss, we iterate over the set as well. The only difference here from a cache hit is that for each way, we read and write auxiliary registers, such as the victim register or the keys register. Reading and updating the whole set is done at the end of this process.
\end{itemize}
In summary, for most policies, the number of operations on a miss is at most
$1\cdot (TCAM\ +\ register\_read\ +\ register\_write)\ +\ 2k\cdot(register\_read\ +\ register\_write)$
and
$1\cdot (TCAM\ +\ register\_read\ +\ register\_write)$
operations for a cache hit.
Hence, hits are handled quickly in $O(1)$ while the latency for handling a miss depends on $k$, due to the need to scan the entire set to find the appropriate victim.
Luckily, as we show in Section~\ref{Experimental Results}, even when $k = 8$ we obtain comparable results to fully associative caches.
In some specific policies there may be a few additional operations.
E.g., in Hyperbolic cache we use an extra resister for log lookup, and when using the TinyLFU filter we access another register for the filter.

\subsection{Limitations and Tradeoffs}
As described before, there are some restrictions in \PFOUR{} and its corresponding BMv2 target simulator. 
Even though we can overcome some of the limitations (like the absence of loops) to implement \PKache{}, there are some restrictions that we cannot. 
One of the restrictions is that TCAM has a limit on the length of a mask, which is $2048$. 
Thus, if our key is for example $32$ bits, it means that $k$ is limited to $64$ entries.

We can, however, alter $d$ to enlarge our \emph{cache\_size} even with this limitation on $k$. 
There is a trade-off between large $k$ and $d$. 
On the one hand, a large $k$ value means that we can do a single comparison on a wide number of elements to check whether the key is inside the cache or not.
Also, a large $k$ value brings the hit ratio closer to that of full associativity, although with very rapidly diminishing returns.
However, bigger $d$ means that elements are more distributed throughout the cache.
This reduces the chance that multiple hot items will reside in the same set, thereby improving the cache hit ratio.
In the case of a multi region cache, we can use different $d$ and different $k$ values for each of the \WindowCache{} and the~\MainCache{}. 

\section{Experimental Results} 
\label{Experimental Results}

As mentioned in the introduction, \emph{hit ratio} is defined as the ratio between the number of accesses that are found in the cache vs. the total number of accesses.
Obviously, the main goal of caches is to obtain high hit ratios, as this is how they improve the overall system's performance.
Hence, we apply this metric to evaluate the effectiveness of our solution.
In particular, we study the impact on the hit ratio due to the choice of cache policy (among the policies we implemented), different configurations (for both \MainCache{} and \WindowCache{}), varying $k$ values and total cache size.
For our tests, we used both synthetic and real traces with varying levels of frequency distribution skewness. 
These traces include:
\begin{description}
\item[Multi3~\cite{jiang2002lirs}:] A trace obtained by executing four workloads (cpp, gnuplot, glimpse, postgres) concurrently.
\item[Sprite~\cite{jiang2002lirs}:] A file system trace of the Sprite network which contains
requests to a file server from client workstations over a two-day period.
\item [Wikipedia(1192951682)~\cite{wikipedia}:] A trace that contains a part of 10\% of all user requests issued to Wikipedia (in all languages) during the period of three months at the end of 2007. This specific dataset contains $4.7$M items.
\item [OTLP~\cite{megiddo2003arc}:] A file system trace of an OLTP server. Note that in a typical OLTP server, most operations are performed on objects already in memory and thus have no direct reflection on disk accesses.
\item [Gradle~\cite{CaffeineProject}:] A trace from the Gradle distributed build cache that holds the compiled output so that subsequent builds on different machines can fetch the results instead of building new ones. It is very recency biased. 
\ifdefined\MWVERSION
\else
Since it operates as a build cache, edited files need to be compiled, therefore they are accessed. But after some times, they stop being accessed. However, they may remain in the cache in case of frequency based policies, thereby polluting the cache. 
\fi
\item[Zipf $s$~\cite{zipf}:] An artificially generated datasets of Zipf distributions with parameter ${s \in \{0.6, 0.99, 1.5\}}$. Each dataset contains $1$M items. 
In a Zipf distribution, the frequency of any element is inversely proportional to its rank in the frequency table. 
Formally, we denote \emph{N} the number of elements, \emph{l} their rank and \emph{s} the value of the exponent characterizing the distribution (skewness). 
The frequency of a flow with rank $l$ is calculated as follow: $f(N, l, s) = \frac{1/l^s}{\sum^{N}_{n=1} (1/n^s)}$. 

\ifdefined\MWVERSION
\else
As the skewness factor increases, the frequency difference between one rank to its next gets larger. 
In highly skewed traces, there are a few very dominant heavy items, so it is easy to track them. 
When the trace is mildly skewed (heavy-tailed trace), like Zipf0.6, the differences between closely ranked items are minor.
Hence, it is much harder to identify the real heavy items that are worth keeping in the cache.
\fi
\end{description}

\subsection{Single Region Cache}
In this section, we examine \PKache's operational envelop over four different aspects:
\begin{itemize}
    \item The impact of Integer factor when used in Hyperbolic \PKache{} on the hit rate.
    \item Comparison between \PKache{} and a pure \emph{Python} implementation (i.e, that is not bounded by \PFOUR's limitations).
    \item The impact of $k$ given fixed \emph{cache\_size} ($k \cdot d = \mathrm{const}$).
    \item The impact of enlarging the cache size.
\end{itemize}

\subsubsection{\PKache{} Hyperbolic Cache Integer Factor}
\begin{table*}
\begin{center}
\scriptsize
\begin{tabular}{|c|c|c|c|c|c|c|c|c|}
\hline
\textbf{} & \textbf{OLTP} & \textbf{Multi3} &\textbf{Wikipedia} & \textbf{Zipf0.6} & \textbf{Zipf0.99} & \textbf{Zipf1.5} &\textbf{Sprite} &\textbf{Gradle} \\ \hline
\textbf{\PKache{} (Integer factor=0.1)} & 10.0945\%  & 8.1525\% & 23.1059\% & 27.1986\% & 60.5630\% & 91.8984\%  & 27.0456\% & 44.1559\% \\ \hline
\textbf{\PKache{} (Integer factor=1)} & 10.1417\%  & 8.1984\% & 23.2129\% & 27.3134\% & 60.9874\% & 92.0045\%  & 27.1676\% & 44.3599\% \\ \hline
\textbf{\PKache{} (Integer factor=10)} & 10.1656\%  & 8.2713\% & 23.3567\% & 27.8849\% & 61.2774\% & 92.1265\%  & 27.3798\% & 44.8592\% \\ \hline
\textbf{\PKache{} (Integer factor=100)} & 10.1734\%  & 8.2790\% & 23.9045\% & 27.3330\% & 61.32711\% & 92.9856\%  & 27.3801\% & 44.8599\% \\ \hline
\textbf{\PKache{} (Integer factor=1000)} & 10.1820\%  & 8.2799\% & 23.9215\% & 27.3430\% & 61.4387\% & 93.0027\%  & 27.3848\% & 44.8603\% \\ \hline
\end{tabular}
\end{center}
\normalsize
\caption{\PKache{} Hyperbolic with $k_m=8, d_m=16$}
\label{tab:hyper}
\vspace{-1mm}
%
\scriptsize
\begin{center}
\begin{tabular}{|c|c|c|c|c|c|c|c|c|}
\hline
\textbf{} & \textbf{OLTP} & \textbf{Multi3} &\textbf{Wikipedia} & \textbf{Zipf0.6} & \textbf{Zipf0.99} &\textbf{Zipf1.5} &\textbf{Sprite} &\textbf{Gradle} \\ \hline
\textbf{\emph{Python} (Fully associative)} & 10.34\% & 8.17\% & 22.21\% & 26.38\% & 61.33\% & 93.33\% & 27.36\% & 43.66\% \\ \hline
\textbf{\emph{Python}} & 10.29\% & 8.61\% & 21.56\% & 26.3\% & 61.38\% & 93.43\% & 27.46\% & 44.2\% \\ \hline
\textbf{\PKache{}} & 10.29\% & 8.61\% &  21.56\% & 26.3\% & 61.38\% & 93.43\% & 27.46\% & 44.2\% \\ \hline
\end{tabular}
\normalsize
\end{center}
\caption{\PKache{} and \emph{Python} LRU with $k_m=8, d_m=16$}
\label{tab:pythonlru}
\vspace{-1mm}
%
\begin{center}
\scriptsize
\begin{tabular}{|c|c|c|c|c|c|c|c|c|}
\hline
\textbf{} & \textbf{OLTP} & \textbf{Multi3} &\textbf{Wikipedia} & \textbf{Zipf0.6} & \textbf{Zipf0.99} &\textbf{Zipf1.5} &\textbf{Sprite} &\textbf{Gradle} \\ \hline
\textbf{\emph{Python} (Fully associative)} & 4.05\% & 11.04\% & 33.17\% & 34.51\% & 66.84\% & 94\% & 13.25\% & 4.42\% \\ \hline
\textbf{\emph{Python}} & 5.83\% & 14.69\% & 29.76\% & 33.95\% & 66.32\% & 93.91\% & 16.7\% & 11.96\% \\ \hline
\textbf{\PKache{}} & 5.27\% & 12.02\% & 25.98\% & 31.74\% & 64.97\% & 93.28\% & 16.01\% & 9.99\% \\ \hline
\end{tabular}
\normalsize
\end{center}
\caption{\PKache{} and \emph{Python} LFU with $k_m=8, d_m=16$ -- the reason why for recency biased traces (OLTP, Multi3, Sprite, Gradle) fully associative LFU is worse than limited associativity is explained in the text}
\label{tab:pythonlfu}
\vspace{-1mm}
\end{table*}

The first set of experiments studies the impact of the Integer factor on the \emph{hit ratio}.
The results 
\ifdefined\MWVERSION
\else
of varying the Integer factor from $0.1$ to $1000$ with the different traces
\fi
are listed in Table~\ref{tab:hyper}.
As shown, increasing the Integer factor improves the hit ratio.
Yet, as long as the Integer factor is $10$ or above, the difference is at most $\pm 0.01\%$ in these measurements.
When the Integer factor is $1$ or $0.1$, the difference is larger.
This is because now many items become indistinguishable, which may result in incorrect evictions.
Still, the largest difference we found was below $1.5\%$.
This is expected since in any case, the exact number is only a rough indicator for the probability that an item will be accessed again shortly.
\ifdefined\MWVERSION
\else
In summary, the Integer factor has only a marginal impact on the hit ratio.
\fi

There is a tradeoff between the Integer factor and memory consumption.
When the former is small, the $log_2$ lookup table requires fewer entries since its values are small.
For example, suppose the maximal value is $2048$.
For an Integer factor of $0.1$ we need just $3$ bits to represent the value ($2$) whereas for an Integer factor of $100$, we need $11$ bits to represent the value ($2000$). 
Hence, for Integer factor $0.1$, we need $14$ bits in total ($3$ bits for the log value and $11$ bits for key) multiplied by $2048$ (number of possible values) - $3.5$KB, while for Integer factor of $100$, we need $22$ bits multiplied by $2048$ - $5.6$KB (around $160\%$).
From now on, we fix the Integer factor at~$100$.

\subsubsection{\emph{Python} Implementation vs \PKache{}}


Next, we compare \PKache{} to a pure \emph{Python} implementation, which is not bound by \PFOUR's restricted programming model and is therefore potentially more accurate.
We have implemented in \emph{Python} both a fully associative cache and a $k$-way cache.
Table~\ref{tab:pythonlru} exhibits the results for \PKache{} with the LRU cache replacement policy while Table~\ref{tab:pythonlfu} lists the results for LFU.
As shown, the LRU results are nearly identical in all~implementations.

With LFU the situation is more involved.
First, the results of \PKache{} are often worse than the limited associativity \emph{Python} implementation.
This is because in LFU there may be multiple items whose frequency is the same and hence are considered equal eviction candidates.
In our \emph{Python} implementation, such symmetry is broken by preferring to remove the least recently used items.
\ifdefined\FULLPAPER
This is also what most unrestricted LFU implementations do.
\fi
In contrast, in \PKache{}, elements may change their position within their set during the reduction process.
When we are looking for a victim, we take as a victim the first element with the lowest \SCN{}.
Hence, the symmetry between the same frequency items is broken in a somewhat arbitrary manner.
This is not the case with LRU; since every \SCN{} is unique, there is no symmetry to break.

Even though our implementation for LFU in \PKache{} is restricted, the traces exhibit an expected behavior: Zipf traces, Multi3 and Wikipedia, which are more frequency biased in their nature, perform better with LFU than LRU, while OLTP, Sprite and Gradle results are better with LRU. 
Further, for recency biased traces, fully associative LFU is worse than limited associativity.
This is because for recency biased traces, frequency can in fact be an anti-signal.
Specifically, past frequent items that are no longer being accessed pollute the cache.
In this particular case, limited associativity limits this ``damage''.
This is further discussed in Section~\ref{sec:k-behavior} below.


\subsubsection{Impact of Different $k$ values on \PKache{}}
\label{sec:k-behavior}
\begin{figure*}
\begin{minipage}[b]{.33\textwidth}
\centering
\begin{tikzpicture}[scale=0.60,font=\normalsize]
\begin{axis}[
    xlabel={$k_m$},
    ylabel={Hit rate \%},
    xmin=8, xmax=64,
    ymin=0, ymax=25,
    xtick={8,16,32,64,100},
    ytick={0, 5, 10, 15, 20, 25},
    legend pos=south west,
    ymajorgrids=true,
    grid style=dashed,
]
\addplot[
    color=blue,
    mark=square,
    ]
    coordinates {
    (8,20.82)(16,20.82)(32,20.88)(64,20.85)
    };
    \addlegendentry{FIFO}
    
\addplot[
    color=red,
    mark=o,
    ]
    coordinates {
    (8,22.22)(16,23.44)(32,23.62)(64,23.65)
    };
    \addlegendentry{LRU}
    
\addplot[
    color=green,
    mark=asterisk,
    ]
    coordinates {
    (8,19.31)(16,19.37)(32,19.27)(64,19.35)
    };
    \addlegendentry{LFU}
    
\addplot[
    color=purple,
    mark=triangle,
    ]
    coordinates {
    (8,21.67)(16,21.63)(32,21.72)(64,21.5)
    };
    \addlegendentry{Hyperbolic (Integer factor=100)}

\end{axis}
\end{tikzpicture}
\captionsetup{size=small}
\captionof{figure}{Hit rate for OLTP} \label{fig:singleoltp}
\end{minipage}%
\begin{minipage}[b]{.33\textwidth}
\centering
\begin{tikzpicture}[scale=0.60,font=\normalsize]
\begin{axis}[
    xlabel={$k_m$},
    ylabel={Hit rate \%},
    xmin=8, xmax=64,
    ymin=0, ymax=90,
    xtick={8,16,32,64,100},
    ytick={0, 10, 20, 30, 40, 50, 60, 70, 80, 90},
    legend pos=south west,
    ymajorgrids=true,
    grid style=dashed,
]
\addplot[
    color=blue,
    mark=square,
    ]
    coordinates {
    (8,81.85)(16,81.85)(32,81.73)(64,81.85)
    };
    \addlegendentry{FIFO}
    
\addplot[
    color=red,
    mark=o,
    ]
    coordinates {
    (8,84.17)(16,84.34)(32,84.33)(64,84.55)
    };
    \addlegendentry{LRU}
    
\addplot[
    color=green,
    mark=asterisk,
    ]
    coordinates {
    (8,84.95)(16,85.16)(32,85.22)(64,85.37)
    };
    \addlegendentry{LFU}
    
\addplot[
    color=purple,
    mark=triangle,
    ]
    coordinates {
    (8,83.37)(16,84.54)(32,84.76)(64,84.92)
    };
    \addlegendentry{Hyperbolic (Integer factor=100)}

\end{axis}
\end{tikzpicture}
\captionsetup{size=small}
\captionof{figure}{Hit rate for Zipf0.99} 
\label{fig:singlezipf}
\end{minipage}%
\begin{minipage}[b]{.33\textwidth}
\centering
\begin{tikzpicture}[scale=0.60,font=\normalsize]
\begin{axis}[
    xlabel={$k_m$},
    ylabel={Hit rate \%},
    xmin=8, xmax=64,
    ymin=0, ymax=40,
    xtick={8,16,32,64,100},
    ytick={0, 10, 20, 30, 40, 50},
    legend pos=south west,
    ymajorgrids=true,
    grid style=dashed,
]
\addplot[
    color=blue,
    mark=square,
    ]
    coordinates {
    (8,32.14)(16,32.23)(32,32.24)(64,32.29)
    };
    \addlegendentry{FIFO}
    
\addplot[
    color=red,
    mark=o,
    ]
    coordinates {
    (8,35.99)(16,37.31)(32,37.43)(64,37.45)
    };
    \addlegendentry{LRU}
    
\addplot[
    color=green,
    mark=asterisk,
    ]
    coordinates {
    (8,39.19)(16,39.76)(32,39.55)(64,39.6)
    };
    \addlegendentry{LFU}
    
\addplot[
    color=purple,
    mark=triangle,
    ]
    coordinates {
    (8,35.89)(16,37.67)(32,37.98)(64,38)
    };
    \addlegendentry{Hyperbolic (Integer factor=100)}

\end{axis}
\end{tikzpicture}
\captionsetup{size=small}
\captionof{figure}{Hit rate for Wikipedia} 
\label{fig:singlewiki}
\end{minipage}
\begin{minipage}[b]{.33\textwidth}
\centering
\begin{tikzpicture}[scale=0.60,font=\normalsize]
\begin{axis}[
    xlabel={$k_m$},
    ylabel={Hit rate \%},
    xmin=8, xmax=64,
    ymin=0, ymax=40,
    xtick={8,16,32,64,100},
    ytick={0, 10, 20, 30, 40},
    legend pos=south west,
    scaled ticks=false,
    ymajorgrids=true,
    grid style=dashed,
]
\addplot[
    color=blue,
    mark=square,
    ]
    coordinates {
    (8,24.94)(16,24.94)(32,24.95)(64,25.01)
    };
    \addlegendentry{FIFO}
    
\addplot[
    color=red,
    mark=o,
    ]
    coordinates {
    (8,31.18)(16,31.71)(32,31.84)(64,32.21)
    };
    \addlegendentry{LRU}
    
\addplot[
    color=green,
    mark=asterisk,
    ]
    coordinates {
    (8,33.17)(16,33.69)(32,33.74)(64,33.90)
    };
    \addlegendentry{LFU}
    
\addplot[
    color=purple,
    mark=triangle,
    ]
    coordinates {
    (8,29.84)(16,30.57)(32,30.79)(64,31.05)
    };
    \addlegendentry{Hyperbolic (Integer factor=100)}

\end{axis}
\end{tikzpicture}
\captionsetup{size=small}
\captionof{figure}{Hit rate for Multi3}
\label{fig:singlemulti3}
\end{minipage}%
\begin{minipage}[b]{.33\textwidth}
\centering
\begin{tikzpicture}[scale=0.60,font=\normalsize]
\begin{axis}[
    xlabel={$k_m$},
    ylabel={Hit rate \%},
    xmin=8, xmax=64,
    ymin=0, ymax=80,
    xtick={8,16,32,64,100},
    ytick={0, 20, 40, 60, 80},
    legend pos=south west,
    scaled ticks=false,
    ymajorgrids=true,
    grid style=dashed,
]
\addplot[
    color=blue,
    mark=square,
    ]
    coordinates {
    (8,73.5)(16,73.78)(32,74.17)(64,74.29)
    };
    \addlegendentry{FIFO}
    
\addplot[
    color=red,
    mark=o,
    ]
    coordinates {
    (8,77.56)(16,78.31)(32,78.81)(64,78.9)
    };
    \addlegendentry{LRU}
    
\addplot[
    color=green,
    mark=asterisk,
    ]
    coordinates {
    (8,65.79)(16,65.87)(32,66.55)(64,66.89)
    };
    \addlegendentry{LFU}
    
\addplot[
    color=purple,
    mark=triangle,
    ]
    coordinates {
    (8,77.12)(16,77.32)(32,77.72)(64,78.00)
    };
    \addlegendentry{Hyperbolic (Integer factor=100)}

\end{axis}
\end{tikzpicture}
\captionsetup{size=small}
\captionof{figure}{Hit rate for Sprite}
\label{fig:singlesprite}
\end{minipage}%
\begin{minipage}[b]{.33\textwidth}
\centering
\begin{tikzpicture}[scale=0.60,font=\normalsize]
\begin{axis}[
    xlabel={$k_m$},
    ylabel={Hit rate \%},
    xmin=8, xmax=64,
    ymin=0, ymax=80,
    xtick={8,16,32,64,100},
    ytick={0, 20, 40, 60, 80},
    legend pos=south west,
    scaled ticks=false,
    ymajorgrids=true,
    grid style=dashed,
]
\addplot[
    color=blue,
    mark=square,
    ]
    coordinates {
    (8,66.92)(16,66.80)(32,66.70)(64,66.67)
    };
    \addlegendentry{FIFO}
    
\addplot[
    color=red,
    mark=o,
    ]
    coordinates {
    (8,68.06)(16,67.91)(32,67.80)(64,67.76)
    };
    \addlegendentry{LRU}
    
\addplot[
    color=green,
    mark=asterisk,
    ]
    coordinates {
    (8,32.19)(16,22.49)(32,15.85)(64,11.92)
    };
    \addlegendentry{LFU}
    
\addplot[
    color=purple,
    mark=triangle,
    ]
    coordinates {
    (8,68.58)(16,68.54)(32,68.46)(64,68.29)
    };
    \addlegendentry{Hyperbolic (Integer factor=100)}

\end{axis}
\end{tikzpicture}
\captionsetup{size=small}
\captionof{figure}{Hit rate for Gradle}
\label{fig:singlegradle}
\end{minipage}
\end{figure*}

Next, we explore the impact that the value of $k$ has on measured performance.
We run the OLTP, Zipf0.99, Wikipedia, Multi3, Sprite and Gradle traces with all \PKache{} single policies modes and different $k$ values for the same \emph{cache\_size} of $512$ items ($k \cdot d = 512$) and measured the hit ratio.
As can be seen, in the OLTP trace (Figure~\ref{fig:singleoltp}) the impact for LFU, Hyperbolic, and FIFO is marginal and inconclusive.
For LRU, the hit ratio increases with the value of $k$, but it is also merely $0.5\%$. 

In Zipf0.99 (Figure~\ref{fig:singlezipf}), Wikipedia (Figure~\ref{fig:singlewiki}) and Multi3 (Figure~\ref{fig:singlemulti3}), the impact for LFU, LRU and FIFO is at most $0.5\%$.
Yet, for Hyperbolic, there is an increase of hit ratio with the value of $k$ by at most $1.5\%$ (between $k=8$ and the rest of the $k$'s). 
This is because the Hyperbolic implementation, as mentioned before, is not accurate.
Hence, increasing $k$ may limit the ``damage'' caused by these inaccuracy errors when we potentially replace the ``wrong'' item (there is a much higher chance of an error when the $k$ is small) compared to what an implementation that is not bounded by \PFOUR{} restricted programming model would do.
This is especially true for both Zipf and Wikipedia as they are frequency biased, hence the error in the frequency might play a more significant role.


Gradle's behavior is non-intuitive (Figure~\ref{fig:singlegradle}). 
\ifdefined\FULLPAPER
In this trace, increasing $k$ damages the hit rate (in all policies).
While for FIFO, LRU, and Hyperbolic the decline is marginal, the decline in LFU is very noticeable ($32.19\%$ vs.$11.92\%$). 
\else
In this trace, increasing $k$ damages the hit rate, and is especially noticeable for LFU ($32.19\%$ vs.$11.92\%$).
\fi
When $k$ is small, we have more sets ($d$), so keys are more evenly distributed between the sets.
This reduces the likelihood of multiple hot items residing in the same set.
In contrast, when $k$ is bigger, the cache becomes closer to a fully associative one. 
Frequency is almost an anti-signal for Gradle.
Hence, a fully associative cache with LFU performs the worse in this trace. 

Another observation is that in the recency biased traces, namely OLTP (Figure~\ref{fig:singleoltp}), Sprite (Figure~\ref{fig:singlesprite}) and Gradle (Figure~\ref{fig:singlegradle}), the best policies are those involving recency (either LRU or Hyperbolic). 
\ifdefined\FULLPAPER
In Wikipedia (Figure~\ref{fig:singlewiki}), Multi3 (Figure~\ref{fig:singlemulti3}) and Zipf (Figure~\ref{fig:singlewiki}), on the other hand, the LFU policy is slightly better, which is well in line with the traces' nature (since all of them tend to be more frequently~biased).
\else
In Wikipedia (Figure~\ref{fig:singlewiki}), Multi3 (Figure~\ref{fig:singlemulti3}) and Zipf (Figure~\ref{fig:singlewiki}) the LFU policy is slightly better due to the traces' frequency biased~nature.
\fi

\begin{figure}[t]
\centering
\begin{minipage}[b]{.45\textwidth}
\centering
\begin{tikzpicture}[scale=0.8,font=\normalsize]
\begin{axis}[
    xlabel={$cache\_size$ (notice the logarithmic scale)},
    ylabel={Hit rate \%},
    xmin=7, xmax=11,
    ymin=0, ymax=50,
    xtick={7,8,9,10,11},
    xticklabels={$2^{7}$,$2^{8}$,$2^{9}$,$2^{10}$,$2^{11}$},
    scaled ticks=false,
    ytick={0, 10, 20, 30, 40, 50},
    legend pos=south west,
    ymajorgrids=true,
    grid style=dashed,
]
    
\addplot[
    color=red,
    mark=o,
    ]
    coordinates {
    (7,10.19)(8,16.53)(9, 23.44)(10,32.82)(11,42.39)
    };
    \addlegendentry{LRU}

\end{axis}
\end{tikzpicture}
\captionsetup{size=small}
\captionof{figure}{Hit rate for OLTP trace for varying cache sizes}  \label{fig:oltpcache}
\end{minipage}%
\hspace{15pt}
\begin{minipage}[b]{.45\textwidth}
\centering
\begin{tikzpicture}[scale=0.8,font=\normalsize]
\begin{axis}[
    xlabel={$cache\_size$ (notice the logarithmic scale)},
    ylabel={Hit rate \%},
    xmin=7, xmax=11,
    ymin=0, ymax=50,
    xtick={7,8,9,10,11},
    xticklabels={$2^{7}$,$2^{8}$,$2^{9}$,$2^{10}$,$2^{11}$},
    ytick={0, 10, 20, 30, 40, 50},
    scaled ticks=false,
    legend pos=south west,
    ymajorgrids=true,
    grid style=dashed,
]
    
\addplot[
    color=red,
    mark=o,
    ]
    coordinates {
    (7,8.27)(8,20.00)(9, 30.57)(10,38.83)(11,46.26)
    };
    \addlegendentry{Hyperbolic (Int. factor=100)}

\end{axis}
\end{tikzpicture}
\captionsetup{size=small}
\captionof{figure}{Hit rate for Multi3 trace for varying cache sizes}  \label{fig:multi3cache}
\end{minipage}
\end{figure}

\subsubsection{The impact of enlarging the cache size}
As reported, e.g., in~\cite{kway, AdaptiveTinyLFU}, OLTP can obtain a hit rate of over 40\% when the cache size is around 1700 items. 
We would like to test whether we can achieve these high rates in \PKache{} with a larger cache (the cache size in Figure~\ref{fig:singleoltp} is 512, and the cache size for Table~\ref{tab:pythonlru} and Table~\ref{tab:pythonlfu} is 128). 
To that end, we took $k_m=16$ and measured hit rates while varying cache sizes. 
\ifdefined\MWVERSION
The policy we chose is LRU, since OLTP is a recency biased trace.
\else
The policy we chose is LRU, since OLTP is a recency biased trace, so we wanted to take a policy that has to consider recency in its eviction process.
\fi
Figure~\ref{fig:oltpcache} shows the \emph{hit rate} of \PKache{}. 
As we can see, indeed when we have a larger cache size, the hit rate of \PKache{} reaches 40\% as expected from previous~findings.

As reported in~\cite{jiang2002lirs}, Multi3 can obtain a $40$\% hit rate when the cache size is around 1700 items. 
Again, we would like to verify that \PKache{} can equal these numbers.
For this, we chose Hyperbolic cache even though this is a frequency biased trace. 
The reason we chose this policy and not LFU is that we wish to see if we can get to these levels of hit rate even when we are not using the best policy for this trace, and even if the implementation is not purely accurate due to the restricted model in \PFOUR{}.
Figure~\ref{fig:multi3cache} shows the \emph{hit rate} of \PKache{}. 
Indeed, with a larger cache size, the hit rate of \PKache{} reaches 40\% as in previous~findings.


In summary, our findings echo previously published results for these traces on fully associative caches with similar cache sizes and cache management policies, e.g.,~\cite{einziger2017tinylfu,jiang2002lirs,megiddo2003arc}.
\ifdefined\FULLPAPER
\fi

\begin{table*}[t]
\begin{center}
\scriptsize
\begin{tabular}{|c|c|c|c|c|}
\hline
\textbf{} & \textbf{FIFO\texttimes LRU\texttimes TinyLFU} & \textbf{FIFO\texttimes LFU\texttimes TinyLFU} & \textbf{FIFO\texttimes LRU} & \textbf{LRU\texttimes LRU\texttimes TinyLFU}\\ \hline
\textbf{\emph{Python} (Fully associative)} & 24.20\% & 13.64\% & 18.81\% & 24.86\% \\ \hline
\textbf{\emph{Python}} & 24.44\% & 12.76\% & 22.71\% & 24.63\% \\ \hline
\textbf{\PKache} & 23.51\% & 12.07\% & 22.71\% & 23.74\% \\ \hline
\end{tabular}
\end{center}
\caption{OLTP on multi-region with $k_w=4, k_m=16, d_w=d_m=16$}
\label{tab:oltp}
\vspace{-1mm}
%
\scriptsize
\begin{center}
\begin{tabular}{|c|c|c|c|c|}
\hline
\textbf{} & \textbf{FIFO\texttimes LRU\texttimes TinyLFU} & \textbf{FIFO\texttimes LFU\texttimes TinyLFU} & \textbf{FIFO\texttimes LRU} & \textbf{LRU\texttimes LRU\texttimes TinyLFU}\\ \hline
\textbf{\emph{Python} (Fully associative)} & 78.46\% & 78.61\% & 76.7\% & 78.7\% \\ \hline
\textbf{\emph{Python}} & 78.37\% & 78.9\%  & 76.76\% & 78.43\% \\ \hline
\textbf{\PKache} & 76.32\% & 76.8\%  & 76.76\% & 76.39\% \\ \hline
\end{tabular}
\end{center}
\caption{Zipf0.99 on multi-region with $k_w=4, k_m=16, d_w=d_m=16$}
\label{tab:zipf099}
\vspace{-1mm}
%
\scriptsize
\begin{center}
\begin{tabular}{|c|c|c|c|c|}
\hline
\textbf{} & \textbf{FIFO\texttimes LRU\texttimes TinyLFU} & \textbf{FIFO\texttimes LFU\texttimes TinyLFU} & \textbf{FIFO\texttimes LRU} & \textbf{LRU\texttimes LRU\texttimes TinyLFU}\\ \hline
\textbf{\emph{Python} (Fully associative)} & 35.67\% & 43.91\% & 33.02\% & 35.68\% \\ \hline
\textbf{\emph{Python}} & 35.37\% & 41.26\% & 32.66\% & 35.51\% \\ \hline
\textbf{\PKache} & 33.89\% & 39.5\% & 32.66\% & 34.67\% \\ \hline
\end{tabular}
\end{center}
\caption{Wikipedia on multi-region with $k_w=4, k_m=16, d_w=d_m=16$}
\label{tab:wiki}
\vspace{-1mm}
%
\scriptsize
\begin{center}
\begin{tabular}{|c|c|c|c|c|}
\hline
\textbf{} & \textbf{FIFO\texttimes LRU\texttimes TinyLFU} & \textbf{FIFO\texttimes LFU\texttimes TinyLFU} & \textbf{FIFO\texttimes LRU} & \textbf{LRU\texttimes LRU\texttimes TinyLFU}\\ \hline
\textbf{\emph{Python} (Fully associative)} & 32.29\% & 35.44\% & 31.50\% & 35.29\% \\ \hline
\textbf{\emph{Python}} & 35.01\% & 36.15\% & 31.05\% & 35.47\% \\ \hline
\textbf{\PKache} & 34.07\% & 34.78\% & 31.05\% & 34.89\% \\ \hline
\end{tabular}
\end{center}
\caption{Multi3 on multi-region with $k_w=4, k_m=16, d_w=d_m=16$}
\label{tab:multi}
\vspace{-1mm}
%
\scriptsize
\begin{center}
\begin{tabular}{|c|c|c|c|c|}
\hline
\textbf{} & \textbf{FIFO\texttimes LRU\texttimes TinyLFU} & \textbf{FIFO\texttimes LFU\texttimes TinyLFU} & \textbf{FIFO\texttimes LRU} & \textbf{LRU\texttimes LRU\texttimes TinyLFU}\\ \hline
\textbf{\emph{Python} (Fully associative)} & 58.40\% & 34.31\% & 60.97\% & 59.78\% \\ \hline
\textbf{\emph{Python}} & 59.85\% & 39.01\% & 59.44\% & 59.92\% \\ \hline
\textbf{\PKache} & 59.15\% & 37.57\% & 59.44\% & 59.12\% \\ \hline
\end{tabular}
\end{center}
\caption{Sprite on multi-region with $k_w=4, k_m=16, d_w=d_m=16$}
\label{tab:sprite}
\vspace{-1mm}
%
\scriptsize
\begin{center}
\begin{tabular}{|c|c|c|c|c|}
\hline
\textbf{} & \textbf{FIFO\texttimes LRU\texttimes TinyLFU} & \textbf{FIFO\texttimes LFU\texttimes TinyLFU} & \textbf{FIFO\texttimes LRU} & \textbf{LRU\texttimes LRU\texttimes TinyLFU}\\ \hline
\textbf{\emph{Python} (Fully associative)} & 35.00\% & 21.00\% & 64.34\% & 37.74\% \\ \hline
\textbf{\emph{Python}} & 36.25\% & 27.63\% & 64.25\% & 36.58\% \\ \hline
\textbf{\PKache} & 35.72\% & 25.89\% & 64.25\% & 35.98\% \\ \hline
\end{tabular}
\end{center}
\caption{Gradle on multi-region with $k_w=4, k_m=16, d_w=d_m=16$}
\label{tab:gradle}
\vspace{-1mm}
\end{table*}

\subsection{Multiple Region Cache}

To test the results of multiple region cache deployments, we took the following configuration $k_w=4, k_m=16, d_w=d_m=16$.
We measured the obtained hit ratios with OLTP, Zipf0.99, Wikipedia, Multi3, Sprite and Gradle. We also measured a fully associative cache of the same size that was implemented in \emph{Python} ($k_w=64, k_m=256, d_w=d_m=1$).

For each trace, we measured the following policy combinations: FIFO in window cache and LRU in the main cache (we denote it FIFO\texttimes LRU\texttimes TinyLFU), FIFO in the window cache and LFU in the main (denoted FIFO\texttimes LFU\texttimes TinyLFU), FIFO in the window and LRU in the main without the TinyLFU filter (denoted FIFO\texttimes LRU) and LRU in both the window cache and the main (denoted by LRU\texttimes LRU\texttimes TinyLFU). Note that the latter configuration is the one used in~\cite{einziger2017tinylfu,CaffeineProject}.
The results for OLTP are reported in Table~\ref{tab:oltp}, for Zipf0.99 in Table~\ref{tab:zipf099}, for Wikipedia in Table~\ref{tab:wiki} for Multi3 in Table~\ref{tab:multi}, for Sprite in Table~\ref{tab:sprite} and for Gradle in Table~\ref{tab:gradle}.

As shown, FIFO\texttimes LRU behaves the same between the \emph{Python} implementation and \PKache{} in all traces, for the reasons discussed above. 
When we add a TinyLFU filter, the results are no longer the same. This is due to the de-amortized aging process. 
While the Python implementation for aging frequencies is by dividing all counters by 2 at the same time, in \PKache{} we divide by 2 only some of the counters, but more frequently. 
Hence, the TinyLFU filter may take somewhat different decisions on whether the element from the \WindowCache{} should enter the \MainCache{} or not. 

In general, when the hit rate is higher, the difference between \PKache{} and \emph{Python} is higher.
This makes sense because when the hit ratio is high, the cost of a wrong eviction can be higher, and there is a higher chance that the de-amortization process will have an impact (since not all counters are divided at the same time, the possibility of wrong partial comparison is higher).
Still, in Zipf the difference is just over $2\%$, in OLTP it is less than $1\%$, 
\ifdefined\MWVERSION
and approximately $1.5\%$ for the rest.
\else
and for Wiki, Multi3, Sprite and Gradle is approximately $1.5\%$.
\fi

When comparing the multi-cache region to the single region (Figure~\ref{fig:singleoltp}) in OLTP, even when the \emph{cache\_size} is smaller (for the single trace we took $k_m=16$ and $d_m=32$, \emph{cache\_size} was 512, and now 320), our results for the \{FIFO,LRU\}\texttimes LRU\texttimes TinyLFU are better than the results for just LRU ($23.51\%$, $23.74\%$ vs. $23.44\%$). 
The same is true for Multi3 (Figure~\ref{fig:multi3cache}) ($34.07\%$, $34.89\%$ vs. $31.71\%$). 
With Wikipedia (Figure~\ref{fig:singlewiki}), our FIFO\texttimes LFU \texttimes TinyLFU is very close to the LFU configuration in the single region ($39.5\%$ vs. is $39.76\%$).
Further, for all recency biased traces, (OLTP, Sprite and Gradle), LRU\texttimes LRU\texttimes TinyLFU is better than FIFO\texttimes LRU\texttimes TinyLFU, as expected. 

Interestingly, in almost all traces, FIFO\texttimes LRU obtains lower hit ratios than FIFO\texttimes LRU\texttimes TinyLFU. 
The only exceptions are Gradle and Multi3. 
In Gradle, as we explained earlier, and as can be seen in Figure~\ref{fig:singlegradle} and Table~\ref{tab:gradle}, any consideration of frequency harms the results, so it is not surprising that adding the TinyLFU filer is worsening the hit ratio. 
In Sprite (Table~\ref{tab:sprite}), which is another recency biased trace, the difference is very marginal.
This is in line with the results published in~\cite{einziger2017tinylfu} motivating the use of a two-region configuration with an admission filter between them.


There is an anomaly in Zipf0.99 in which in \emph{Python}, FIFO\texttimes LRU is worse than FIFO\texttimes LRU\texttimes TinyLFU by about 2\%, whereas in \PKache{} there is much smaller margin (0.4\%).
This is explained by the fact that \PKache's TinyLFU filter implementation is not accurate.
Hence, when the hit rate is so large, our mistakes may have a more noticeable impact.
Further, with Multi3 in \emph{Python}, FIFO\texttimes LFU\texttimes TinyLFU is the best (as we can expect), but in \PKache{} it is only the second best. 
This can be explained by the fact that our implementation for LFU is not accurate, so when the trace is frequency biased, our mistakes may have more noticeable impact as well.

\ifdefined\VLDBVERSION
\else
Overall, when comparing our $k$-way cache against the fully associative one, the results are encouraging. 
In quite a few traces and configurations, our \PKache{} beats the fully associative \emph{Python} implementation.
The most noticeable is Gradle (FIFO\texttimes LFU\texttimes TinyLFU), which is not surprising, but also in Sprite (FIFO\texttimes LRU\texttimes TinyLFU) and more. Clearly, there are other traces and configurations where fully associative is better.
\fi

\section{Discussion}
\label{Conclusions and Discussion}

We have presented \PKache, a generic limited associativity design for implementing caching in the data plane.
\PKache{} supports multiple popular caching schemes, both when the entire cache is treated as a single region or when we have multiple regions.
The limited associativity design helps overcome certain \PFOUR{} and PISA limitations, such as lack of iterators, and facilitates bounded fast handling of packets.
We enable instantiating \PKache{} with varying parameters by generating its \PFOUR{} code automatically from our set of \emph{Python} scripts~\cite{PKache-code}.
Let us emphasize that our goal in this work is not to promote any specific cache management policy.
Rather, \PKache{} is intended to be a generic framework that facilitates realizing existing as well as novel policies inside the data plane.

We demonstrated our multi-region capabilities using \TinyLFU, because it is a recent policy that is also very widely adopted~\cite{CaffeineProject,RistrettoProject}, and because its use of an admission filter between the two regions adds an extra challenge.
Yet, our multi-region support can be applied to other policies involving more than one region such as the seminal 2Q~\cite{2Q} policy and the recent FRD~\cite{park2017frd} scheme.
Here, the filter would simply always evaluate to true.
An interesting challenge left for future work is how to support multi-region caches in which the relative size of each region changes at runtime, such as ARC~\cite{megiddo2003arc} and Adaptive W-TinyLFU~\cite{AdaptiveTinyLFU}.

\ifdefined\FULLPAPER
One immediate extension of our work is to replace the TinyLFU filter implementation to work with CM-Sketch~\cite{cormode2005improved} instead of the explicit counting structure.
An immediate improvement of our work
is to augment the LFU behavior of \PKache{} to break the symmetry between same frequency objects based on their recency rather than arbitrarily.
A simple method to realize this is to store in the \SCN{} field both the frequency counter and access timestamp, and compare objects' utility based on both measures (first frequency and then recency).
\fi

\ifdefined\MWVERSION
\else
In this work, we ignored potential differences in object sizes.
This is because in many storage systems, cached objects have the same sizes or nearly the same size, often a page or block size.
Further, when there are significant size variations, the overall cache area is often partitioned into slabs, where each slab holds objects of similar size and is treated as an independent size oblivious cache, e.g.,~\cite{redis}.
\ifdefined\FULLPAPER
Yet, there are situations, e.g., CDNs, where size-aware cache management can out-perform a slabbing based solution~\cite{LHD,AdaptSize,GDSF}.
\fi
A slabbing based solution can easily fit limited associativity designs by adding the object's size range to the hashing function that maps an object to its respective set.
We leave combining a limited associativity design with a non-slabbed size-aware cache~management policy like~\cite{LHD,AdaptSize,GDSF} for future work.
\fi

\printbibliography

\end{document}